          [2004/06/22 1.2 (KOF); 2001/04/25 1.1 (PWD)]

\documentclass[a4paper,twocolumn]{Gaia2004} % European paper size
\usepackage{times}      % for font
\usepackage{epsfig}     % for figure inclusion
\usepackage{natbib}     % for bibliography

\def\av{A$_{\rm V}$}
\def\feh{[Fe/H]}
\def\logg{$\log g$}
\def\teff{${\rm T}_{\rm eff}$}
\def\cw{$c$}
\def\hwidth{$b$}
\def\itime{$t$}
\def\deg{$^{\circ}$}

\title{Design of the Gaia photometric systems
for stellar parametrization using a population-based optimizer}

\author{C.A.L.\ Bailer-Jones}
\affil{Max-Planck-Institut f\"ur Astronomie, K\"onigstuhl 17, 69117
Heidelberg, Germany}

\bibpunct{(}{)}{;}{a}{}{,}  % to set bibliography punctuation to A&A style

\begin{document}

\keywords{photometric systems -- stellar
parameters -- optimization -- 
evolutionary algorithms -- Gaia}

\maketitle

\begin{abstract}
  
Large, deep surveys must typically rely on multiband photometry rather
than spectroscopy for
determining the astrophysical properties (APs) of stars. Yet designing
an optimal photometric system for a wide range of objects is complex,
because it must trade off conflicting scientific requirements.  I
present a new method for designing photometric systems and apply it to
the Gaia Galactic survey satellite, which will observe one billion
stars brighter than V=20. The principle is to optimally sample stellar
spectra in order to best determine the APs (e.g.\ \teff, [Fe/H],
interstellar extinction).  By considering a filter system as a
set of free parameters (central wavelengths, FWHM etc.), it
may be designed by optimizing a figure-of-merit (FoM) with respect to
these parameters.  The FoM is a measure of how well the filter system
can vectorially `separate' stars in the data space to
avoid AP degeneracies.  The resulting systems show some interesting
features, in particular broad, overlapping filters, which may
be desirable from a multivariate classification perspective.  These
systems are competitive with others proposed for Gaia.

\end{abstract}

\section{Introduction}
Surveys of large numbers of objects will often be forced to employ
photometry rather than spectroscopy due to confusion and SNR
considerations. Given well defined scientific goals, the designer must
decide how many filters to use, with what kind of profiles, where to
locate them in the spectrum and how much integration time to assign to
each. This is usually achieved via a manual inspection of typical
target spectra.  But if the survey is intended to establish multiple
astrophysical parameters (APs) across a large and varied population of
objects, then this method is unlikely to be very efficient or even
successful. Even if a reasonable filter system could be constructed in
this way, we would not know whether a better filter system exists
subject to the same constraints.  In this contribution I outline a
more systematic approach to designing filter systems by optimizing
a figure-of-merit (FoM) with respect to a parametrization of the
filter system. The FoM is a measure of how well the filter system can
`separate' stars with different APs.  This separation is vectorial in
nature, in the sense that the local directions of AP variance are
preferably mutually orthogonal to avoid AP degeneracy. I apply this
model, HFD (Heuristic Filter Design), to the design of the photometric
system for the Gaia mission.  HFD and results
of its application are described in more detail in Bailer-Jones
(2004a,b).

\section{Evolutionary Algorithms}
Evolutionary Algorithms (EAs) are stochastic population-based
optimizers which use principles of evolutionary biology to perform a
directed search (e.g.\ Goldberg 1989). A population of individuals
(candidate solutions) is evolved over many generations (iterations)
making use of specific {\it genetic operators} to modify the genes
(parameters) of the individuals. The goal is to locate the maximum of
some {\it fitness} function (figure-of-merit).  As a population-based
method, it takes advantage of evolutionary behaviour (breeding,
natural selection, maintenance of diversity etc.) to perform more
efficient searches than single solution methods.  The algorithm works
iteratively: starting from some initial population, the fitness of
each individual is calculated. A selection operator is then applied,
in which individuals reproduce with a probability proportional to
their fitness, i.e.\ fitter individuals produce more offspring.  These
offspring are produced by applying small random changes to the parents
({\it mutation}). This forms the next population and the procedure is
iterated, as shown in Fig.~1.

\begin{figure}[ht]
\begin{center}
\epsfig{file=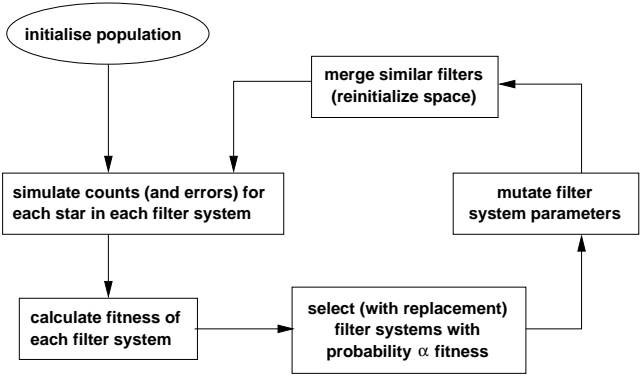, width=8.5cm, angle=0}
\end{center}
\caption{The iterative loop (one iteration) of HFD.}
\label{fig1}
\end{figure}

\section{Method}
A filter system consists of $I$ filters. Each filter is parametrized
with three parameters: the central wavelength, \cw, the half-width at
half maximum (HWHM), \hwidth, and the fractional integration time,
\itime, per star allocated to this filter. (The total
integration time per star over the whole mission is fixed.)  The
profile shape of the filter is fixed. The optimization is performed
with respect to these $3I$ free parameters.

The purpose of a filter system is to enable us to determine multiple
stellar astrophysical parameters (APs), such as \teff, \feh\ etc.  To
achieve this, the filter system must maximally separate stars with
different APs. A suitable measure of this is defined as follows.

\begin{figure}[ht]
\begin{center}
\epsfig{file=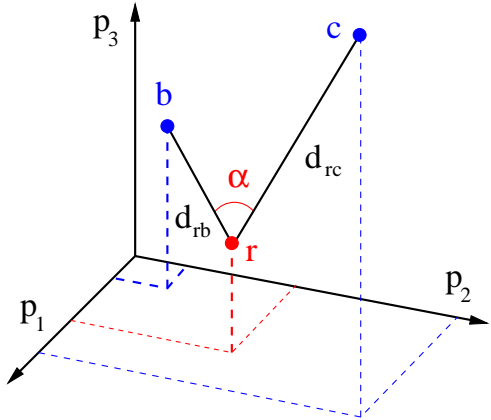, width=6.5cm, angle=0}
\end{center}
\caption{The fitness measures the separation between different types
of stars as well as the angle between the directions in which the
astrophysical parameters vary.}
\label{fig2}
\end{figure}

Given a fixed grid of stellar spectra, we synthesize the photometry
for each of these stars in a given filter system and thus determine
where they lie in the $I$-dimensional data space. At any point in this
space, each AP will vary in a certain direction (the {\em principal
direction}), and at a certain rate, the (scalar) {\em AP-gradient}.
Fig.~2 shows an example for $I=3$.  A good filter system will have
large AP-gradients, i.e.\ a given pair of stars will be separated by a
large amount (in proportion to their AP difference).  This is
necessary but not sufficient: We must also ensure that the {\it
angles} between the principal directions are as near to 90\deg\ as
possible, to minimize the degeneracy between APs. The fitness of a
filter system is a function which reflects these (for details see the box {\it
Fitness: the maths}).

It is important to realise that the fitness is calculated using a
specified grid of spectra. This grid (and the APs which vary in it)
represents the types of objects we want the filter system to
parametrize. In the example which follows, the grid shows variance in
\teff, \logg, \feh\ and \av\ over relatively large ranges.

\begin{figure}[ht]
\begin{center}
\epsfig{file=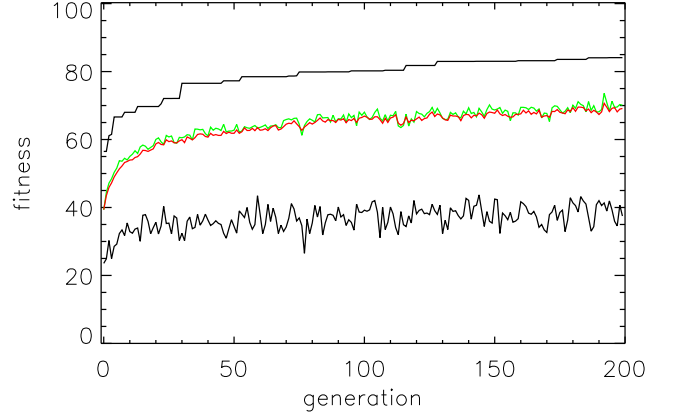, width=6.0cm, angle=90}
\end{center}
\caption{Fitness evolution of a population, showing (from top to bottom)
the maximum, mean, median and minimum fitnesses.}
\label{fig3}
\end{figure}

\begin{figure*}[!ht]
\begin{center}
\epsfig{file=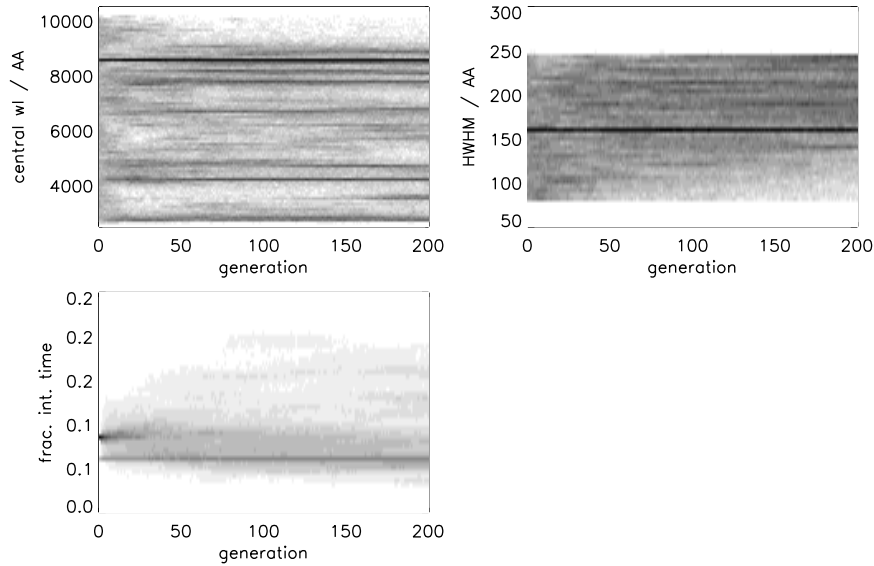, width=8cm, angle=90}
\end{center}
\caption{Evolution of the filter system parameters for systems with 12
filters, shown for all 200 filter systems (`individuals') in the
population.  At each generation all $12\times200$ corresponding parameters are
plotted (as a density grey scale, i.e.\ darker is a higher density).}
\label{fig4}
\end{figure*}

\begin{table}
\begin{center}
\begin{tabular}{|p{7.5cm}|}
\hline
{\bf Fitness: the maths}

\vspace*{2ex}
The spectral grid shows variance in $J$ astrophysical parameters (APs).
For each star, $r$, in the grid, we find its $J$ nearest
neighbours, each of which differs from $r$ in only one of the $J$
APs. The relevant `distance' between $r$ and that neighbour differing
in AP $j$ (call it $n_j$), is the {\it AP-gradient} and is defined as
\[
h_{r,n_j} = \frac{ d_{r,n_j} }{ | \Delta \phi_{r,n_j} |}
\]
where $d_{r,n_j}$ is the Euclidean distance between $r$ and $n_j$ in
SNR units and $\Delta \phi_{r,n_j}$ is their difference in AP $j$.
Clearly, the larger $h$ the better we have separated $r$ and $n_j$.
To minimise the degeneracy between the principal
directions to these $J$ neighbours we want angle
$\alpha$ in Fig.~2 to be as close to 90\deg\ as possible
for all neighbour pairs. Combining these measures, we see that a
useful figure-of-merit of separation is
\[
x_{r,j,j'} = h_{r,n_j} \, h_{r,n_{j'}} \, {\sin \alpha_{r,j,j'}}
\]
where $j$ and $j'$ label those neighbours which differ from $r$ in APs
$j$ and $j'$ respectively. (This is the magnitude
of the cross product between the two vectors.) For $J$ APs we have
$J(J-1)/2$ pairs of neighbours and thus $J(J-1)/2$ terms like that above,
Summing these over all stars in the grid gives the
final fitness, which is to be maximized
\[
F = \sum_{j,j' \neq j} \sum_r x_{r,j,j'} \ \ \ .
\]
(The actual fitness function is a slight modification of this. See
Bailer-Jones (2004) for details.)\\ \hline
\end{tabular}
\end{center}
\end{table}

\section{Results}
HFD is applied to design the Gaia photometric system.  The
optimization is performed within specified limits on the wavelength
coverage of any filter plus maximum and minimum values of the HWHM of
the filters As an example, an optimization is performed with 12
filters. A population size of 200 is used and is evolved for 200
generations.  Multiple runs are performed with different initial
(random) populations.  An example of the evolution of the fitnesses in
the population is shown in Fig.~3; the corresponding evolution of the
filter system parameters is shown in Fig.~4. From each of many runs
such as these, the best filter system is selected. Examples are shown
in Fig.~5.

\begin{figure*}[!ht]
\begin{center}
\epsfig{file=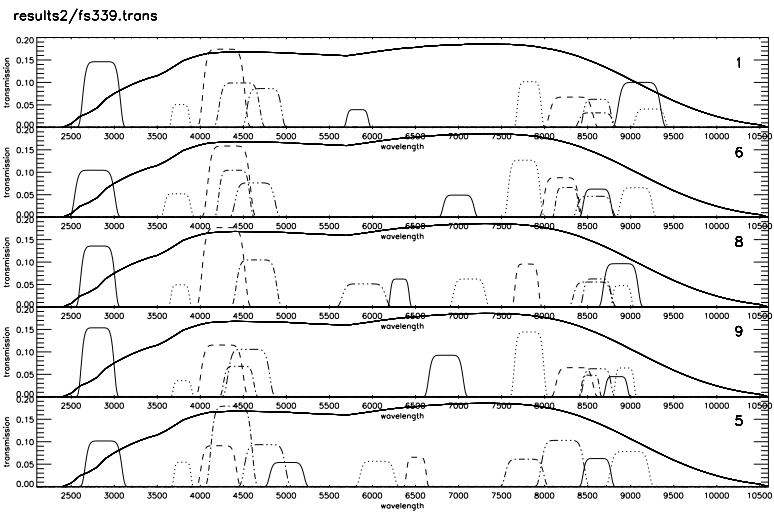, width=7cm, angle=90}
\end{center}
\caption{Five HFD-optimized filter systems. HFD also optimizes the
integration time allocated to each filter. To show this the filter
transmission curves have been multiplied by their fractional
integration times (with the result that the intercepts of the profiles
are not accurately depicted). The true filters all have a peak
transmission of 0.9.}
\label{fig5}
\end{figure*}

\section{Conclusions}
Various filter systems have been produced with HFD for Gaia. The best
have a performance competitive with or better than
conventionally-designed systems. One proposed for Gaia is shown in
Fig.~6. Many optimization with different parameter settings have been carried out. General conclusions from this work are as follows:
\begin{itemize}
\item{HFD often shows a preference for broad, overlapping filters.
These may have advantages over traditional systems as they may make a
better use of a high dimensional data space. (AP signatures are
coherent over a wide wavelength range so broad filters can in principle still be sensitive to AP variations -- see Fig.~7.)}
\item{There is a trade-off between broad and narrow filters: narrower
filters degrade the `scalar' separation (as they achieve lower SNR)
but improve the `vector' separation (better at isolating the effects
of APs).}
\item{Some filters are placed very consistently, especially at the very
blue and red ends. The wavelength space is not uniformly populated
with filters, e.g.\ there is often a crowding around 7500--9000\,\AA.}
\item{HFD shows a good ability to converge on a common filter system,
but as the number of filters increases, there is less
convergence. Although the different optimal filter systems show quite
different properties, they achieve similar performance. In other words
there are many different `good' filter systems (roughly equal local
optima).}
\end{itemize}
HFD is a systematic and powerful approach to designing filter systems
for surveys. Its principles are generic and it may be applied to
many other survey projects.  

\begin{figure}[!ht]
\begin{center}
\epsfig{file=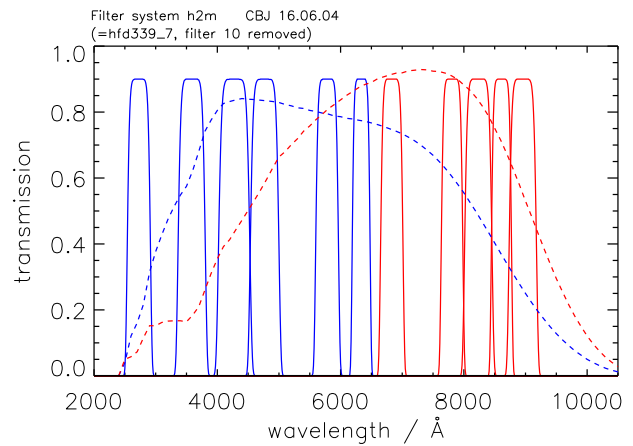, width=6cm, angle=90}
\end{center}
\caption{An 11-band system designed by HFD for stellar
parametrization, and proposed for Gaia.}
\label{fig6}
\end{figure}

\begin{figure*}[!ht]
\begin{center}
\epsfig{file=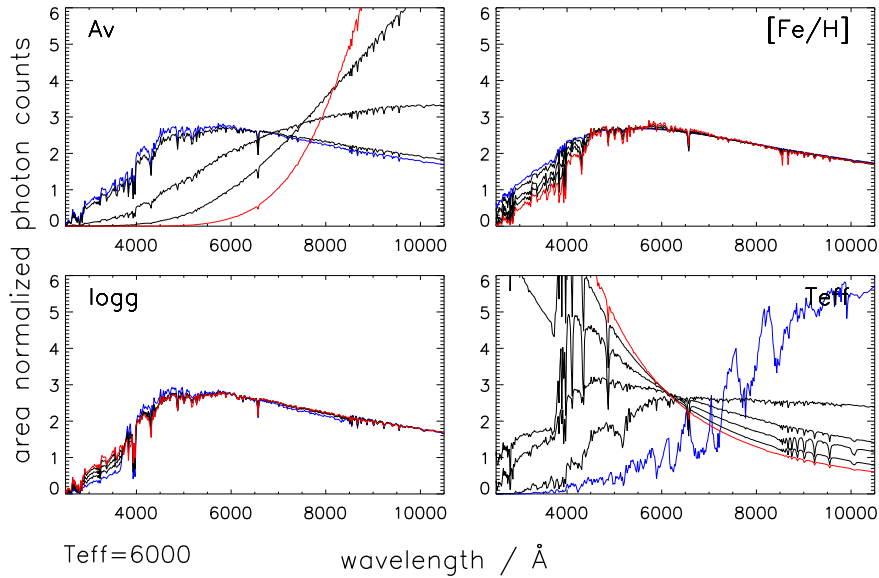, width=8cm, angle=90}
\end{center}
\caption{Examples of the effects that varying the four astrophysical
parameters (APs) have on the spectra. (Each panel shows the effect of
varying just the AP labelled.) The effect of APs is often coherent
over over a wide wavelength range, meaning that broad (and
overlapping) filters can be used for stellar parametrization, as found
by HFD.}
\label{fig7}
\end{figure*}

%\section*{Acknowledgments}


\begin{thebibliography}{}
\bibitem[2004a]{bailer-jones04a}
Bailer-Jones, C.A.L., 2004a, A\&A 419, 385

\bibitem[2004b]{bailer-jones04b}
Bailer-Jones, C.A.L., Gaia working group report, GAIA-CBJ-016

\bibitem[1989]{goldberg89}
Goldberg, D.E., 1989, {\it Genetic algorithms in search, optimization, and
machine learning}, Addison-Wesley, Reading, MA

 \end{thebibliography}
\end{document}